\documentclass[preprint,preprintnumbers,amsmath,amssymb, superscriptaddress]{revtex4}

\usepackage{graphicx}
\usepackage{dcolumn}
\usepackage{bm}
\begin{document}


\title{Momentum dependence of superconducting gap, strong-coupling dispersion kink, and tightly bound Cooper pairs in the high-$T_c$ (Sr,Ba)$_{1-x}$(K,Na)$_x$Fe$_2$As$_2$ superconductors}


\author{L. Wray}
\affiliation{Joseph Henry Laboratories of Physics, Department of Physics, Princeton
University, Princeton, NJ 08544, USA}
\author{D. Qian}
\affiliation{Joseph Henry Laboratories of Physics, Department of Physics, Princeton
University, Princeton, NJ 08544, USA}
\author{D. Hsieh}
\affiliation{Joseph Henry Laboratories of Physics, Department of Physics, Princeton
University, Princeton, NJ 08544, USA}
\author{Y. Xia}
\affiliation{Joseph Henry Laboratories of Physics, Department of Physics, Princeton
University, Princeton, NJ 08544, USA}
\author{L. Li}
\affiliation{Joseph Henry Laboratories of Physics,  Department of Physics, Princeton
University, Princeton, NJ 08544, USA}
\author{J.G. Checkelsky}
\affiliation{Joseph Henry Laboratories of Physics,  Department of Physics, Princeton
University, Princeton, NJ 08544, USA}
\author{A. Pasupathy}
\affiliation{Joseph Henry Laboratories of Physics,  Department of Physics, Princeton
University, Princeton, NJ 08544, USA}
\author{K.K. Gomes}
\affiliation{Joseph Henry Laboratories of Physics,  Department of Physics, Princeton
University, Princeton, NJ 08544, USA}
\author{C.V. Parker}
\affiliation{Joseph Henry Laboratories of Physics,  Department of Physics, Princeton
University, Princeton, NJ 08544, USA}
\author{A.V. Fedorov}
\affiliation{Lawrence Berkeley National Laboratory, Advanced Light
Source, Berkeley, CA 94305,
USA}
\author{G.F. Chen}
\affiliation{Beijing National Laboratory for Condensed Matter Physics, Institute of Physics, Chinese Academy of Sciences, Beijing 100080, P.R. China}
\author{J.L. Luo}
\affiliation{Beijing National Laboratory for Condensed Matter Physics, Institute of Physics, Chinese Academy of Sciences, Beijing 100080, P.R. China}
\author{A. Yazdani}
\affiliation{Joseph Henry Laboratories of Physics,  Department of Physics, Princeton
University, Princeton, NJ 08544, USA}
\author{N.P. Ong}
\affiliation{Joseph Henry Laboratories of Physics,  Department of Physics, Princeton
University, Princeton, NJ 08544, USA}
\author{N.L. Wang}
\affiliation{Beijing National Laboratory for Condensed Matter Physics, Institute of Physics, Chinese Academy of Sciences, Beijing 100080, P.R. China}
\author{M.Z. Hasan}
\email [To whom correspondence should be addressed: ]  {mzhasan@Princeton.EDU}
\affiliation{Joseph Henry Laboratories of Physics,  Department of Physics, Princeton
University, Princeton, NJ 08544, USA} \affiliation{Princeton Center
for Complex Materials, Princeton University, Princeton, NJ 08544,
USA}


\date{14$^{th}$ August, 2008}


\begin{abstract}
We present a systematic angle-resolved photoemission spectroscopic
study of the high-T$c$ superconductor class
(Sr/Ba)$_{1-x}$(K/Na)$_x$Fe$_2$As$_2$. By utilizing a
photon-energy-modulation contrast and scattering geometry we report
the Fermi surface and the momentum dependence of the superconducting
gap, $\Delta(\overrightarrow{k})$. A prominent quasiparticle
dispersion kink reflecting strong scattering processes is observed
in a binding-energy range of 25-55 meV in the superconducting state,
and the coherence length or the extent of the Cooper pair wave
function is found to be about 20 $\AA$, which is uncharacteristic of
a superconducting phase realized by the BCS-phonon-retardation
mechanism. The observed 40$\pm$15 meV kink likely reflects
contributions from the frustrated spin excitations in a J$_1$-J$_2$
magnetic background and scattering from the soft phonons. Results
taken collectively provide direct clues to the nature of the pairing
potential including an internal phase-shift factor in the
superconducting order parameter which leads to a Brillouin zone node
in a strong-coupling setting.

\end{abstract}

\maketitle

\begin{figure}[t]
\includegraphics[width=9.2cm]{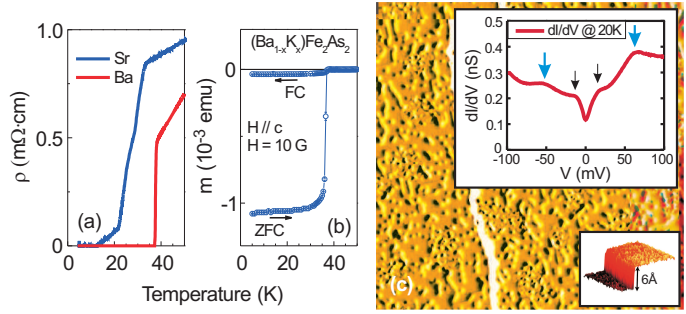}
\caption{\textbf{Phase transition, magnetization, and STM
characterization}. (a-b) Bulk $T_c$ of crystalline $(Ba,K)Fe_2As_2$
and $(Sr,K)Fe_2As_2$ was determined based on the resistivity and
magnetization profiles. $(Ba,K)Fe_2As_2$ samples exhibited $T_c$ =
37K and $\delta$$T_c$ $\sim$ 1K whereas $(Sr,K)Fe_2As_2$ samples
exhibited a broad ($\sim$ 10K) transition with a $T_c$ $\sim$ 26K.
(c) Surface quality was studied by atomic-resolution STM which
revealed a high degree of flatness and confirmed the suitability for
spectroscopic measurements. The derivative of an STM image is shown
which was taken on a 500$\times$500$\AA^2$ patch. The inset shows
STM data in the superconducting state with a gap of
2$\Delta$$\thickapprox$30 meV and kink structures around 40-50 meV
loss-energies.}
\end{figure}

The recent discovery of superconductivity ($T_c$ up to 55K) in
iron-based layered compounds promises a new route to high
temperature superconductivity \cite{kami, gfchen, cruz}. This is
quite remarkable in the view that the $T_c$ in the pnictides is
already larger than that observed in the single-layer cuprates.
Preliminary studies suggest that the superconducting state in these
materials competes with a magnetically ordered state, and the proper
description of the ordered state lies somewhere in between a strong
correlation mediated local moment magnetism and quasi-itineracy with
stripe-like frustration \cite{cruz, singh, kuroki, mazin, yao,
eremin, bernevig, kot, phonon_unlikely, ma, tesan, yildirim,
nematic}. This calls for a microscopic investigation of pair
formation and related electron dynamics in these superconductors.
Angle-resolved photoemission spectroscopy (ARPES) is a powerful tool
for investigating the microscopic electronic behavior of layered
superconductors \cite{arpes, kaminski}. In this work we report
electronic structure results focusing on the details of the
low-lying quasiparticle dynamics on very high quality ($\delta$$T_c$
$\lesssim$ 1K and surface-RMS $\sim$ 1$\AA$) \textit{single domain}
single crystal samples, which allow us to gain insight into
connections between the superconductivity and magnetism. We observe
that the electrons are strongly scattered by collective processes
around the 15 to 50 meV binding energy range depending on the Fermi
surface sheet while a magnitude-oscillating gap structure persists
nearly-along the SDW wave vector of the parent compound. We also
show that a Cooper pair in this superconductor is very tightly bound
($\lesssim$ 4a$_o$). Our overall results can be self-consistently
interpreted in a phase-shifting order parameter scenario.

\begin{figure*}[t]
\includegraphics[width=16cm]{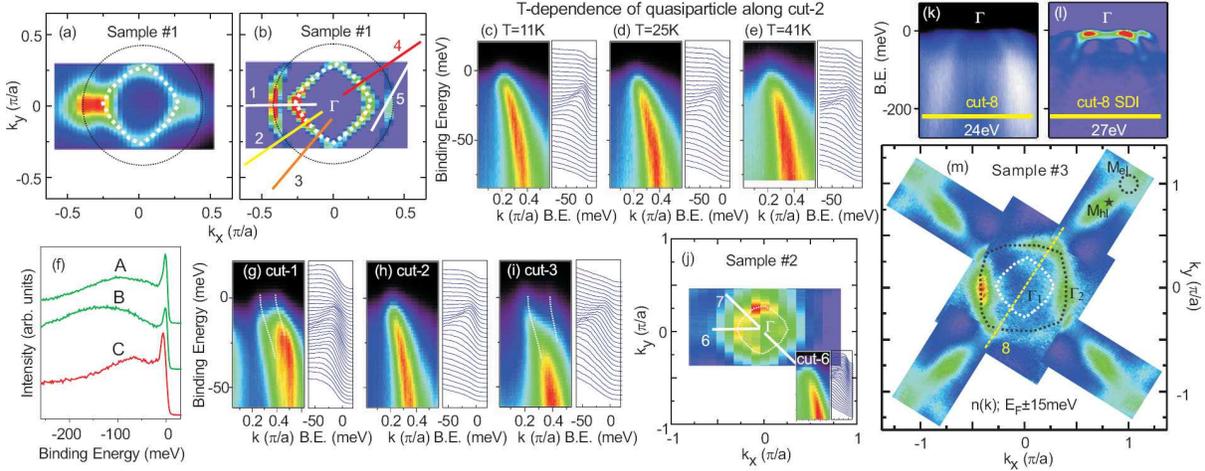}

\caption{\textbf{Fermi surface and quasiparticle behavior:} (a)
ARPES intensity integrated within 15 meV of Fermi level in
$(Ba,K)Fe_2As_2$. (b) Second-derivative image approximation of the
Fermi surface topology around the $\Gamma$-point. (c-e)
Quasiparticle dispersion along cut-2 and its temperature evolution.
(f) High-resolution fine-step binding energy scans shown for some
selected k-space points (A, B) near the [1,0] and [1,1] axes on the
outer FS surrounding the $\Gamma$ point, and (C) on a separate FS
close to the M-point. (g-i) Quasiparticle intensity profiles along
k-space cuts 1 to 3. The k-space cut-2 strongly suppresses the outer
FS and provides a clear spectroscopic look at the quasiparticle that
forms the innermost FS. Because of the spectral clarity this
quasiparticle can be studied in quantitative detail. (j) Fermi
surface image ($\pm$15meV) taken on $(Sr,K)Fe_2As_2$. (k,l) Wide
k-range coarse-step scans are shown which were used for locating the
Fermi crossings. (m) ARPES intensity map within 15 meV of Fermi
energy over the complete BZ measured with a photon energy of 18 eV.
Hole- ($\Gamma_1$, $\Gamma_2$, M$_{hl}$) and electron-like
(M$_{el}$) Fermi sheets are labeled.}
\end{figure*}

ARPES measurements were performed using 18 to 60 eV photons with
better than 8 to 15 meV energy resolution respectively and overall
angular resolution better than 1\% of the Brillouin zone. Most of
the data were taken at the Advanced Light Source beamline 12.0.1 and
a limited data set was taken at SSRL beamline 5-4, using a Scienta
analyzer with chamber pressures lower than 5x10$^{-11}$ torr.
Linearly polarized photons were used for all the study. The angle
between the $\overrightarrow{E}$-field of the incident light and the
normal direction of the cleaved surface was set to about 45 degrees
(at 12.0.1). Single crystalline samples of
Ba$_{1-x}$K$_x$Fe$_2$As$_2$ ($T_c$=37K), Sr$_{1-x}$K$_x$Fe$_2$As$_2$
($T_c$=26K) and Sr$_{1-x}$Na$_x$Fe$_2$As$_2$ ($T_c$=36K) were used
for this systematic and class-independent study of the kink
phenomena. Cleaving the samples in situ at 15K resulted in shiny
flat surfaces. Cleavage properties were thoroughly studied and
characterized by atomic resolution STM measurements and the surface
was found to be flat, with an RMS deviation of 1\AA (Fig.1(c)) and
rarely observed steps of size 6$\AA$. Sample batches with
$\delta$$T_c$ $\sim$ 1K and smooth STM images were selected for UHV
cleaves in the ARPES studies here. The utilization of unique
scattering geometries coupling with specific photon energy contrasts
(18$\pm$2 eV vs. 40$\pm$2 eV) allowed us to selectively suppress one
of the FS sheets so that the other can be studied in details.

\begin{figure}
\includegraphics[width=6.5cm]{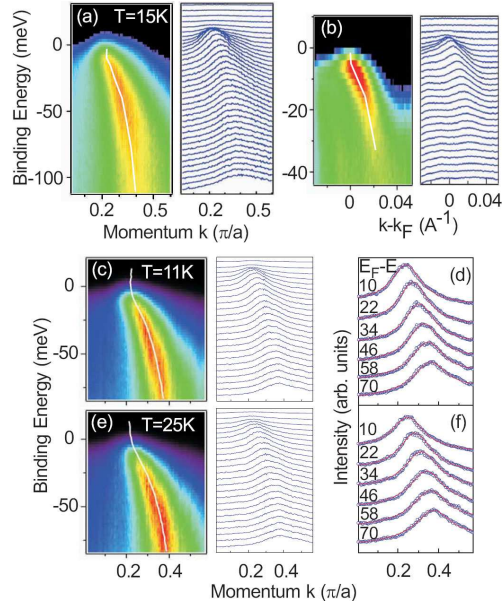}
\caption{\label{fig:Sb_Fig3} (a) Quasiparticle band dispersion along
cut-4 at 15K. The k-space cut-4 is approximately along the
$Q_{AF}$-vector of the undoped compound, as defined in Fig.2(b). (b)
Band dispersion for the $\Gamma_2$ quasiparticle along cut-5 shows a
small kink near 18meV. (c,e) Close-up view of dispersion along cut-4
near the 40meV kink at 11K and 25K. The MDCs can be fitted by
Lorentzians with linear backgrounds. (d),(f) Quasiparticle
lineshapes for panels (c,e) are presented at selected energies
(10-70 meV).}
\end{figure}

Quasiparticle behavior around the $\Gamma$-FS sheets is shown in
Figure 2. Two square-like FS sheets were clearly resolved near the
center of the BZ, labeled $\Gamma_1$ and $\Gamma_2$ in Fig.2(m). An
azimuthal variation of ARPES intensity around the FS pockets was
observed, and found to be most pronounced at 40eV incident energy
for the scattering geometry described above. A comparison of
quasiparticle dispersion measured along the various
$\overrightarrow{k}$-space cuts suggests that looking roughly along
the cuts 30 to 40-degrees to the $\Gamma$ to $(\pi, 0)$-line
provides a clear spectroscopic view of the quasiparticle dispersion
and lineshape behavior on the inner-FS. This is also the cut that is
nearly parallel to the SDW vector (undoped compound). A bend in
dispersion could be observed in the cut-2 data which is not resolved
in cut-1 or 3 data due to the spectral overlap with the outer-FS
(two bands).

A closer look at the quasiparticle dispersion behavior is presented
in Figure 3. A bend in dispersion is evident in the momentum
distribution curves (MDC) taken on a crossing near the $\Gamma_1$-FS
(cut-4). Each MDC could be fitted with a single Lorentzian over a
wide binding energy range and, as in the raw data sets, the fitted
peak positions trace a kink around 40$\pm$15 meV. This is further
seen by examining the peak position of the real part of the
self-energy (Fig.4). Although it is less clear, the MDC width
plotted as a function of the electron binding energy is found to
exhibit a drop below 35 meV which is consistent with a kink in a
nearby binding energy as seen in the raw data. At temperatures above
$T_c$ the kink shifts to somewhat lower energies. As the temperature
is raised further the MDCs are broadened making its identification
or analytic extraction from our experimental data difficult. In the
MDC widths an increase is observed at very low energies which is due
to some residual signal from the tail of the quasiparticles from the
outer FS. Our STM data in Fig.1(c) also exhibit a satellite
structure around 40-50 meV loss-energy range (with respect to the
quasiparticle peak position) consistent with the observed ARPES
kink. Assuming that the kink reflects coupling to some bosonic-like
modes one can estimate the coupling strength:
$\lambda'_{eff}$$\gtrsim$ (0.7/0.45 - 1)$\sim$ 0.6. This coupling is
about a factor of two to three larger than the electron-phonon
coupling ($\lambda_{ph}$$\sim$ 0.2) calculated for the Fe-As phonons
near 20-40 meV \cite{phonon_unlikely, kot}. A careful look at the
outer central FS ($\Gamma_2$ band, cut-5) also reveals a kink around
18 $\pm$ 5 meV. This kink is revealed when the band associated with
the inner-FS sheet is suppressed by a choice of incident photon
energy (18 eV).

\begin{figure}
\includegraphics[width=9cm]{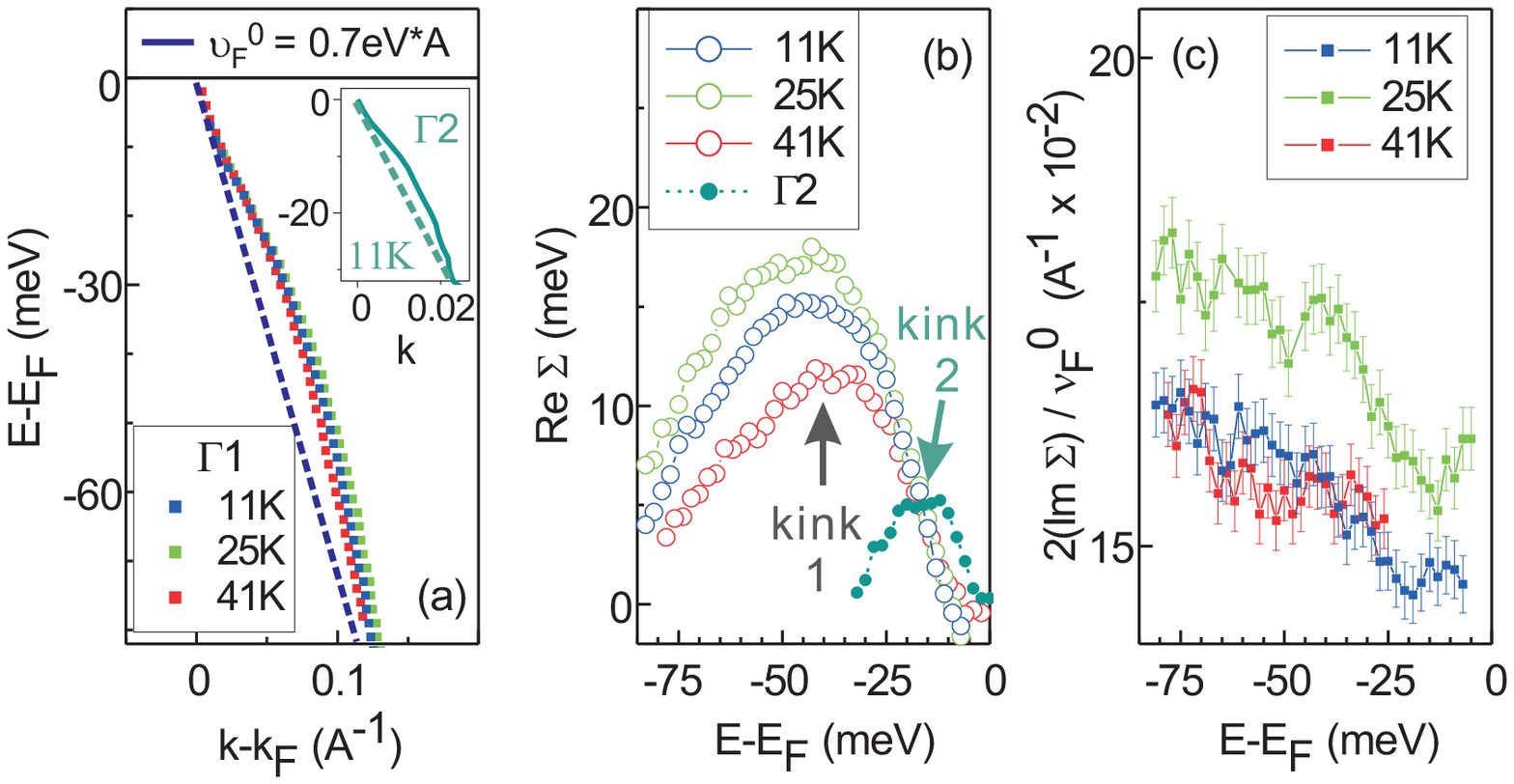}
\caption{\label{fig:Sb_Fig3} (a) By tracing the peak positions,
quasiparticle band dispersion is plotted at T=11K, 25K and 41K. At
all temperatures, the dispersion curve shows a ``kink"-like feature
at 40$\pm$10 meV. The gray dashed line illustrates the ``bare" band
used for the self-energy estimation. The smaller kink at 18$\pm$5meV
on the $\Gamma_2$ band is shown in the inset. (b) The real part of
self energy is obtained by subtracting the gray dashed line from the
experimental band dispersion. Peak position is used to define the
``kink" position. (c) Fitted MDC width as a function of binding
energy for quasiparticles on $\Gamma_1$. A small offset in the T=25K
intrinsic width is due to temperature dependent shifting of beam
position on the sample.}
\end{figure}

If the kink phenomenon is intimately related to the pairing
potential one might expect some inter-correlation between the kink
energies and the superconducting gaps at different FS sheets. In
order to investigate this aspect we present the gap evolution data
taken in these kink-exhibiting samples in Figure 5. The opening of
the superconducting gap is viewed upon symmetrization (see
Ref-\cite{norman} for methods) and a gap magnitude of about 13$\pm$2
meV is quite evident at low temperatures, in agreement with ref.
\cite{dinggap}. This value is consistent with the spatially-averaged
gap ($<$ 15 meV) we have obtained with STM (see Fig.1(c)) on the
same batch of samples and these observations (ARPES in Fig.5 and STM
in Fig.1) confirm that the kinks survive the superconducting gap
formation.
Although our FS topology and an intrinsic fluctuation regime differs
from that in Ref-\cite{kaminski}. Evidently, a more complex gap and
a different FS topology (M-pockets) have been realized in our
$(Ba/Sr,K/Na)Fe_2As_2$ series than that in the NdFeAsO series
\cite{kaminski}. The fine details here, made possible via selective
study of the bands, allow us to look for correlations between the
gap and the kink. The observed kink energies do (Fig.3) seem to
scale (40 meV and 18 meV) with the superconducting gap energies (13
meV and 6 meV) on the two central Fermi surfaces (Fig.4). The
coupled cross-sections of bands near the M-point (M$_{hl}$ and
M$_{el}$ in Fig.2(m)) for modulated incident photon energy make a
systematic study of kinks in that region of momentum space
difficult. We note that our particular gap structure is consistent
with an order-parameter that takes, qualitatively, the form of
$\Delta_o$$cos(k_x)cos(k_y)$ within the plane (see fit in Fig.5(e)).
Although fine details of the gap anisotropy are not resolved due to
lack of resolution, the gap on the $\Gamma_2$ FS is smaller than the
$\Gamma_1$ gap by a $cos(k_x)cos(k_y)$ factor, e.g. along the
$\hat{x}$ axis:
cos(.40$\pi$/$\sqrt{2}$)$^2$/cos(.27$\pi$/$\sqrt{2}$)$^2$=.58$\sim$$\Delta(\Gamma_2)/\Delta(\Gamma_1)$=7meV/14meV.
The large gap ratio, $\frac{2\Delta}{k_BT_C}$=8 for the innermost
gap of 13meV, is a clear sign of strong coupling. The systematic
oscillation pattern and FS topology in our data suggest that the
``node" of the cos$\times$cos order parameter lies in between the
$\Gamma_2$ band and the corner pockets, but it clearly lies outside
(Fig.5) the Fermi contour. Yet, we caution that our ARPES data do
not rule out the possibility of an out-of-plane ($k_z$) node in the
order parameter, therefore the case for completely nodeless
superconductivity remains open. Existence of such a node may explain
the in-gap $T^3$ behavior of NMR data \cite{nmr}, and could
potentially be established by mapping the ARPES gap distribution
over a wide range of incident energies.


A strong-coupling kink phenomenology is observed in the electron
dynamics of high $T_c$ cuprates which occurs around 60$\pm$20 meV,
as observed by ARPES and STM, and is often attributed to phonons or
magnetism or polarons with $\lambda'_{eff}$$\sim$ 1 to 1.5
\cite{arpes}. In cuprates the superexchange coupling is on the order
of 130 meV, whereas the optical phonons are in the range of 40 to 80
meV overlapping with the kink energy. In the pnictides, although a
$T_c$ value of 37K is not outside the phonon-induced strong-coupling
pairing regime, the vibrational modes of the the FeAs plane are
rather soft ($\leq$ 35 meV) making electron-phonon interaction
\cite{phonon_unlikely, kot} an unlikely source of the major part of
the quasiparticle's self-energy beyond 40 meV, considering the
observed coupling $\lambda'_{eff}$$\gtrsim$0.6 for the FeAs
compounds. The parent compounds of superconducting FeAs exhibit a
robust SDW groundstate \cite{cruz} due to a
$\overrightarrow{Q}$=($\pi$, $\pi$) inter-band instability or due to
the interaction of quasi-localized moments and the short range SDW
order seems to survive well into the superconducting doping regime
\cite{phd}. The doping evolution of the Fermi surface lacks robust
nesting conditions for purely band-magnetism to be operative at
these \textit{high dopings} and the relevant magnetism here likely
comes from the local exchange energy scales in a doping induced
frustrated background. Therefore, quite naturally, strong spin
fluctuations in the presence of electron-electron interaction are
important contributors to the self-energy at high dopings. In
accounting for the parent SDW groundstates of these materials the
known values of $J_1$ and $J_2$ are on the order of 20 to 50 meV
\cite{ma, yildirim}. In an itinerant picture, there exists a Stoner
continuum whose energy scales are parameterized by the exchanges
whereas in a local picture, $J_1$ and $J_2$ reflect Fe-Fe and
Fe-As-Fe superexchange paths and the groundstate is a highly
frustrated doped Heisenberg magnet \cite{ma}. The proper description
of the experimentally observed magnetism lies somewhere in between.
In our photoemission process, removal of an electron from the
crystal excites the modes the electron is coupled to, so the
observed quasiparticle breaks the locally frustrated magnetic bonds
(near-neighbor spin correlation) associated with energy-costs
parameterized by $J_1$ and $J_2$ which then contributes an energy
scale on the order of ($J_1 + 2J_2$)/2 $\lesssim$ 50 meV in the
self-energy of the \textit{doped} system. Since this scale is large
it is expected that our 40 meV kink would survive above $T_c$ which
is consistent with our observation. Despite the high signal-to-noise
quality of our data, it is difficult to draw an intimate connection
or relation [kink(18 meV)$\approx$gap(6 meV)+spin-mode(14 meV)]
between the low-energy kink and spin-mode (e.g. magnon)
\cite{spinmode} without a full phase-diagram study.

\begin{figure*}[t]
\includegraphics[width=17cm]{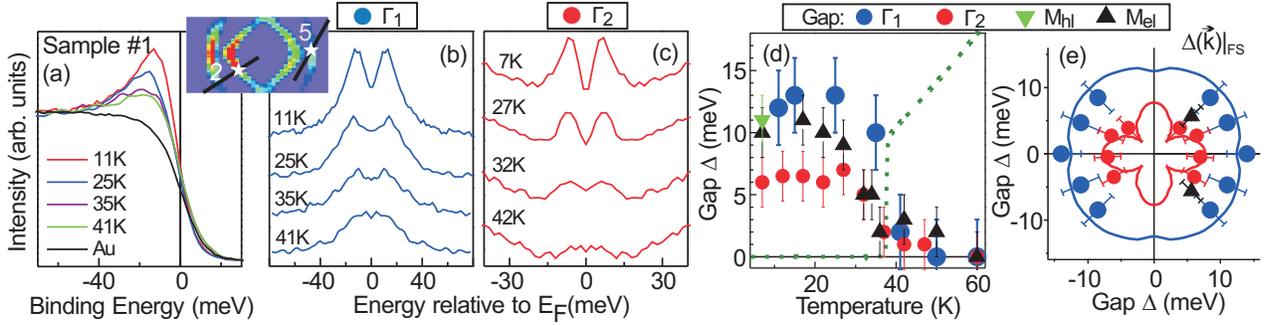}
\caption{\label{fig:Sb_Fig2} \textbf{Momentum and Fermi-surface
dependence of superconducting gap:} (a) temperature dependence of
quasiparticles (cut-2) near the Fermi level through the
superconducting transition. Below $T_c$ samples exhibit
coherence-peak-like behavior similar to what is observed in some
cuprates. Temperature dependence of the gap at the k-space location
of (b) the inner-most ($\Gamma_1$) FS and (c) the outer central
($\Gamma_2$) FS are estimated by symmetrization around $E_F$. (d)
The temperature dependence of the gaps measured at different FS
locations ($\Gamma_1$-FS, blue; $\Gamma_2$-FS, red; $M_{hl}$, green;
$M_{el}$, black) are plotted along with the bulk resistivity curve
(dotted green). A fluctuation regime above $T_c$ is observed. (e)
The azimuthal k-dependences of gaps, $\Delta(k)$, are shown for
different FS sheet locations on a polar plot. Blue and Red solid
lies are contours of the cos$\times$cos function for the Fermi
surface outlined in Fig.2(m). }
\end{figure*}

Our high-resolution (Fig.3) measurements allow us to estimate the
Fermi velocity of the normal state to be about 0.7 eV$\cdot\AA$.
Using the observed superconducting gap (ARPES or STM data in Fig.1)
we can estimate the average size of the Cooper pair wavefunction
$\xi= \frac{\hbar v_F}{\pi\Delta}$ by invoking the uncertainty
relation \cite{tinkham}. Taking average $v_F$(Fig.3 and 2) $\sim$
0.7$\pm$0.1 eV$\cdot\AA$ and a gap (Fig.4) value of $\Delta$ $\sim$
13$\pm$2 meV, this gives $\xi$ $\lesssim$ 20 $\AA$. This value is
remarkably consistent with the high magnitude of H$c_2$ ($\sim$70T)
\cite{hc2} reported in these same materials. The ARPES based Cooper
pair scale and unusually high H$c_2$ clearly suggest that the Cooper
pairs in this class of FeAs superconductors are tightly bound which
is in contrast to the point-contact Andreev spectroscopy results on
Sm-based FeAs superconductors exhibiting a conventional BCS ratio
\cite{jhu}. The agreement between ARPES, bulk H$c_2$ and the bulk
resistivity profile (Fig.5(d)) provides further support for our
identification of the superconducting gap and its
bulk-representative value through our surface-sensitive measurements
(STM and ARPES) here. This also confirms that the ARPES gaps are not
the SDW gaps as theoretically claimed by some authors. More
importantly, such a small Cooper pair size scale ($\sim$ 4$a_o$) is
not known in any phonon-based BCS superconductor \cite{kivelson} but
has only been observed in unconventional strongly correlated
superconductors. Our observed value is much smaller than that in the
s-wave BCS-phonon superconductors such as MgB$_2$ \cite{kivelson}.
In fact, a combination of small Cooper pair size, oscillating but
in-plane nodeless gap function is qualitatively consistent with an
unconventional $\Delta_o$$cos(k_x)cos(k_y)$(in the unfolded BZ with
one iron atom per unit cell)-type or s$_{x^2y^2}$ or s$_{\pm}$ wave
states \cite{kot, kuroki, mazin, yao, eremin, bernevig, tesan} since
such an order parameter (its Fourier transform) has a
nearest-neighbor (NN) or next-NN structure in real space and thus a
reduction of the Coulomb interaction within the pair is naturally
possible, so the electrons can come closer to each other leading to
a short coherence scale. In cuprates, pairing electrons come close
to each other, and the short coherence length is achieved by
introducing a node in the order parameter (\textit{d}-wave) leading
to a reduction of Coulomb interaction within the pair. This is often
the only choice in a single band correlated system (cuprates or
organics). In pnictides, multiband structure can accommodate a phase
change without the need for introducing a ``node" \cite{coscosnode}
on the FS, therefore an isotropic gap and a short pairing scale can
co-exist with a phase shifted order-parameter structure with a BZ
node. Our data suggest that the BZ node lies in between the
$\Gamma_2$ and the corner FS locations along the magnetic wave
vector. Our data also suggest that at higher dopings, (possibly
beyond x=0.4), this BZ node will intersect the sample FS and a
\textit{nodal} superconducting state will be realized. While the
observation of the strong-coupling kink ($\sim$ 40 meV) is an
important first step, its detailed quantitative interpretation will
require complete phase diagram study once single-crystals become
available also at higher dopings.

In summary, we have presented a Fermi surface and momentum
dependence of the superconducting gap study of high-$T_c$
superconductor class (Sr/Ba)$_{1-x}$K$_x$Fe$_2$As$_2$. Our
systematic spectroscopic data suggest an unusually small dimension
of the Cooper pair, kink phenomena (seen both in ARPES and STM
around 40 meV in our data here), and an oscillating gap function,
all of which collectively point to an unconventional pairing
potential. We have presented arguments that in the presence of
magnetism, the observed short pairing scale and a nearly-isotropic
in-plane gap can be self-consistently realized if the order
parameter contains a non-trivial internal $\pi$ phase-shift factor.




\begin{thebibliography}{h}
\bibitem{coscosnode} The cos$\times$cos or s$_{\pm}$ gap does have a node in it but the node does not intersect the Fermi surface as we measure here (Figure 4). However, it is clear that nearest or next nearest neighbour structure of the gap lowers the kinetic energy as opposed to a constant on-site s-wave gap.
\end{thebibliography}
\end{document}